\documentclass[12pt, offprint]{aastex}

\slugcomment{Not to appear in Nonlearned J., 45.}

\shorttitle{Five-minute oscillation power within magnetic elements}
\shortauthors{Jain et al.}

\begin{document}

\title{Five-minute oscillation power within magnetic elements in the solar atmosphere}


\author{Rekha Jain, Andrew Gascoyne}
\affil{School of Mathematics and Statistics, University of Sheffield,
    UK, S3 7RH}
\email{R.Jain@sheffield.ac.uk}

\author{Bradley W. Hindman \& Benjamin Greer}
\affil{JILA, University of Colorado, Boulder, CO~80309-0440, USA}

\begin{abstract}
 
It has long been known that magnetic plage and sunspots are regions in which the power of acoustic waves is reduced within the photospheric layers. Recent observations now suggest that this suppression of power extends into the low chromosphere and is also present in small magnetic elements far from active regions. In this paper we investigate the observed power supression in plage and magnetic elements, by modelling each as a collection of vertically aligned magnetic fibrils and presuming that the velocity within each fibril is the response to buffeting by incident $p$ modes in the surrounding field-free atmosphere. We restrict our attention to modeling observations made near solar disk center, where the line-of-sight velocity is nearly vertical and hence, only the longitudinal component of the motion within the fibril contributes. Therefore, we only consider the excitation of axisymmetric sausage waves and ignore kink oscillations as their motions are primarily horizontal. We compare the vertical motion within the fibril with the vertical motion of the incident $p$ mode by constructing the ratio of their powers. In agreement with observational measurements we find that the total power is suppressed within strong magnetic elements for frequencies below the acoustic cut-off frequency. We also find that the magnitude of the power deficit increases with the height above the photosphere at which the measurement is made. Further, we argue that the area of the solar disk over which the power suppression extends increases as a function of height.

\end{abstract}


\keywords{Sun: MHD --- magnetic fields : waves, Helioseismology}

\section{Introduction}

There is clear observational evidence that strong magnetic structures such as sunspots and plages scatter acoustic waves ($p$-mode oscillations) and modify their propagation. In particular, the measured Doppler power of $p$ modes with frequencies below the acoustic cut-off frequency (i.e., $\nu \approx 5$ mHz) is reduced by about 20--30$\%$ in regions with a strong magnetic field when compared with their weakly magnetised surroundings (e.g., Brown et al. 1992 and references therein; Hindman and Brown 1998; Jain and Haber 2002; Schunker and Braun 2011). The power suppression has been observed in maps of the acoustic power that have been obtained using a variety of different spectral lines with formation heights spread throughout the upper photosphere. Moreover, the spatial extent of the region of power suppression grows as the height increases (Moretti et al. 2007). Possible physical scenarios that could lead to such observed power suppression were suggested in Jain et al. (1996). In particular, they suggest that a magnetic field shortens the attenuation length (or the skin depth) of the $p$-mode eigenfunction in the upper atmosphere where the wave is evanescent.  Such a mechanism would reduce the observed power amplitude without a significant change in the energy carried by the mode.

Recently, Chitta et al. (2012) reported similar power suppression in small-scale magnetic elements with magnetic field strengths $|B| <$ 500 G, both near and far from sunspots. The observed similarities between plage and individual magnetic elements suggests that irrespective of the size of the magnetic regions, the physical mechanism that is responsible for the observed reduction of acoustic power is the same. Thus, it is unlikely that the collective effect of tightly packed magnetic concentrations (as is typical of plage) is responsible. The small scale magnetic structures reported by Chitta et al. (2012) probably consist of several thin magnetic fibrils, each with a different magnetic-field strength with some reaching values of 1--2 kG. Although these intense flux tubes occupy only a small fraction of the quiet Sun photosphere, they are major contributors to the total magnetic flux seen in magnetograms. Such small and intense flux tubes penetrate below the solar surface and generally have strong vertical magnetic fields in the low to mid-photosphere, fan out with height such that near the limb the direction of field appears inclined from vertical (see also Stenflo, 2013).

The interaction of trapped solar $p$ modes with vertically-oriented, thin, magnetic flux tubes has been studied previously by theoretical means (Bogdan et al 1996; Hindman and Jain 2008; Jain et al. 2009; Hindman and Jain 2012). These studies examined many aspects of the excitation of longitudinal (sausage) and transverse (kink) magnetohydrodynamic (MHD) tube waves on such fibrils through buffeting of the tube by the ambient acoustic wave field. However, the atmospheric models employed by all of these studies were truncated for technical reasons such that only layers below the photosphere were considered. Unfortunately, this means that the models failed to extend to heights in the atmosphere where the spectral lines used in observations are typically formed. As a result the models could not be used to predict observed power and thus failed to address the physical mechanism leading to the observed power suppression.

In this paper, we extend these theoretical models by appending an isothermal atmosphere above the subphotospheric model of the solar interior. This isothermal atmsophere is designed to represent the Sun's upper atmosphere and allows magnetically induced changes in the wave function to be studied as a function of height throughout the region where observational spectral lines are formed. Our goal is to model the observed power suppression in both small magnetic elements and in extended regions of plage by comparing the vertical velocity within the magnetic element with the acoustic velocity in the surrounding nonmagnetized atmosphere. In Section 2, we describe the details of the model including the governing equations. The theoretically calculated power ratio and its dependence on flux tube parameters are investigated in Section 3. We discuss these results and their implications for the observed $p$-mode power ratios in Section 4.

\section{The Model}

The observed velocity within a single magnetic element is expected to be a combination of sausage and kink waves, both excited by external forcing by the incident acoustic wave field. Here, we attempt to model only those observations that are made near the center of the solar disk, where the line-of-sight velocity is purely the vertical component. For a vertically aligned tube, the kink wave has only horizontal motion and may be safely ignored. Thus, we only consider the excitation of sausage waves in the following.

As shown in Figure 1, we consider a vertically aligned, thin, magnetic-flux tube embedded within an unmagnetised, gravitationally stratified medium. We denote the equilibrium quantities inside and outside the magnetic flux tube by subscripts $i$ and $e$, respectively. This vertical magnetic fibril is sufficiently thin that it cannot support lateral variations in the magnetic-field strength, $B_i(z)$, gas pressure $p_i(z)$, or mass density $\rho_i(z)$. Further, we assume that the tube is in thermal equilibrium with its surroundings such that the magnetic pressure and gas pressure within the tube vary with the same scale height as the external gas pressure, $p_e(z)$. Since the internal magnetic pressure and gas pressure have the same functional form, their ratio $\beta$, is a constant parameter. We choose to model the solar interior as a neutrally stable polytrope which extends to a height of $z = -z_{0}$. An isothermal upper atmosphere is appended smoothly above the polytrope, and the interface between the two regions is placed at a height such that the temperature of the upper atmosphere corresponds to the Sun's temperature minimum, i.e., $T = 4300$ K.

With these assumptions, the equilibrium pressure and density of the external fluid, $p_{e}(z)$ and $\rho_{e}(z)$ respectively, are:

\begin{eqnarray}
p_{e}(z) = \left\{ \begin{array}{r@{\quad\quad}l}
p_{0}\left(-\frac{z}{z_{0}}\right)^{a+1}\ :& z < -z_{0} \\
p_{0}\ {\rm exp}\left[-\frac{(z+z_{0})}{H} \right]\ : & z \ge -z_{0}
\end{array} \right.
\end{eqnarray}
\noindent and
\begin{eqnarray}
\rho_{e}(z) = \left\{ \begin{array}{r@{\quad\quad}l}
\rho_{0}\left(-\frac{z}{z_{0}}\right)^{a}\ :& z < -z_{0} \\
\frac{p_{0}}{g H}\ {\rm exp}\left[-\frac{(z+z_{0})}{H} \right]\ : & z \ge -z_{0}
\end{array} \right.
\end{eqnarray}

\noindent where $a$ is the polytropic index and $p_{0} = gz_{0}\rho_{0}/(a+1)$ is the pressure at the interface (i.e., $z = -z_{0}$). The adiabatic index $\gamma = 1 + a^{-1}$ is assumed to be 5/3 throughout the atmosphere and the interface pressure $p_{0}$ is chosen to coincide with the reference model of Maltby et al. (1986) at the appropriate temperature, $4300$ K. The interface is at $z = - 260$ km with a sound speed value of 7 km s$^{-1}$. The pressure scale height in the isothermal region, $H = z_{0}/(a+1)$, is roughly 100 km. The corresponding sound speed in the flux tube and the surrounding unmagnetised fluid have identical profiles given by:

\begin{eqnarray}
c_{i}^{2} = c_{e}^{2} = \left\{ \begin{array}{r@{\quad\quad}l}
-\frac{gz}{a}\ :& z < -z_{0} \\
\gamma g H : & z \ge -z_{0}
\end{array} \right.\ .
\end{eqnarray}

\noindent Since the total pressure is balanced at the tube boundary i.e. $p_{i}(z) + B_{i}^{2}(z)/(2 \mu_{0}) = p_{e}(z)$, the tube cross section $A(z)$ increases with height.

Note that the photosphere is represented by the layer with $T = 6000$ K which lies within the polytropic portion of the atmosphere where $z_{\rm photo} = -a c_{\rm photo}^{2}/g$ with sound speed, $c_{\rm photo} = 8$ km s$^{-1}$. For the parameters already provided, the photosphere is located at a height of $z_{\rm photo} \approx -360$ km, or $\sim$ 100 km below the height where the polytrope and isothermal atmosphere join.

Within a neutrally stable polytrope, the vertical velocity of an individual $p$ mode can be expressed using Whittaker functions (Bogdan et al. 1996; Hindman and Jain 2008),
\begin{eqnarray}
\label{external}
v_{z,e} &=& {\cal A}_{p} \, \frac{dQ}{dz} \, \exp\left(ik_x x - \omega t\right) \; ,
\end{eqnarray}
\noindent where
\begin{eqnarray}
Q \equiv (-2k_x z)^{-(\mu+1/2)} \, W_{\kappa,\mu}(-2k_x z) \; ,
\end{eqnarray}
\begin{eqnarray}
\kappa \equiv \frac{a\omega^2}{2k_x g} \; ,\hspace{0.5cm}\mu &\equiv& \frac{a-1}{2}.
\end{eqnarray}
 
\noindent Here, $W$ is the Whittaker function that vanishes asymptotically for large argument, ${\cal A}_{p}$ is an arbitrary amplitude, $\omega$ is the temporal frequency, and $k_x$ is the horizontal wavenumber. Within the isothermal atmosphere, the vertical velocity has the well-known exponential form (see for example, Rae and Roberts, 1982):

\begin{eqnarray}
\label{top}
v_{z,e} &=& {\cal A}_{I} \exp\left(\frac{z+z_{0}}{2H}\right)\, {\rm exp}\left[i k_{z, e}(z+z_{0})\right] \exp\left(ik_x x - \omega t\right) \; ,
\end{eqnarray}
\noindent where
\begin{eqnarray}
\label{kayze}
k_{z,e} &=& \sqrt{\left(\frac{\omega^{2} - \omega_{c}^{2}}{c_{e}^{2}}\right) + k_{x}^{2} \left(\frac{N^{2}}{\omega^{2}} - 1 \right)}.
\end{eqnarray}

\noindent Note that the vertical wavenumber, $k_{z,e}$, depends on the acoustic cut-off frequency, $\omega_{c} =\gamma g/(2 c_{e})$ and the Brunt-V\"ais\"al\"a frequency $N = \left(\gamma - 1\right)^{1/2} g/c_{e}$. We will solely focus on frequencies below the acoustic cut-off frequency such that the $p$ modes are evanescent ($k_{z,e}^2 < 0$) in the upper isothermal region. Thus, the $p$ modes decay with height ($v_{z,e}\sim e^{-\alpha_e z}$) with the rate

\begin{equation}
\label{alphae}
	\alpha_{e} = -\frac{1}{2H} + \sqrt{\left(\frac{\omega_c^{2} - \omega^{2}}{c_{e}^{2}}\right) + k_{x}^{2} \left(1- \frac{N^{2}}{\omega^{2}}\right)} \; .
\end{equation}

The wave amplitude ${\cal A}_{I}$ within the isothermal atmosphere is related to the amplitude ${\cal A}_{p}$ and can be determined by applying a continuity condition at the interface between the polytrope and isothermal atmosphere. By further requiring that the Lagrangian pressure perturbation is continuous across the interface, the eigenfrequencies become quantized $\omega_n(k_x)$ with the solutions correspondings to the $f$ and $p$ modes with different vertical orders $n$ (see Gascoyne et al. 2014).

The $f$ mode and the $p$ modes interact with the thin magnetic-flux tube and excite transverse (kink) and longitudinal (sausage) tube waves along the fibril (Bogdan et al. 1996; Hindman and Jain 2008). Since the sausage waves are the only wave mode that contributes to the vertical velocity component (which is the only visible component with Dopplergrams made at disk center), we only consider sausage wave excitation here and use the subscript $\parallel$ to denote them. The vertical velocity associated with the sausage/longitudinal waves in the polytropic atmosphere has been previously derived by several authors. Using Greens function, the solution can be written as

\begin{eqnarray}
\label{Polyp}
v_{\parallel} &=& -\frac{i\pi}{2}\frac{{\cal A}_{p}}{z_{0}}\left\{\psi_{\parallel}(s) \left[\Omega + {\cal{J}}^{*}(s) \right] + \psi_{\parallel}^{*}(s) \left[{\cal I} - {\cal{J}}(s) \right] \right\} \; ,
\end{eqnarray}

\noindent where we make the following series of nested definitions,

\begin{eqnarray}
s &\equiv& -\frac{z}{z_0} \; ,
\\
\psi_\parallel(s) &=& s^{-\mu/2} H_\mu^{(1)}\left(\Theta s^{1/2}\right) \; ,
\\
{\cal J}(s)&=&-\frac{(a+1)(\beta+1)z_0\omega^2}{2g} \int_{1}^{s} (s^\prime)^{\mu} \, \psi_{\parallel}(s^\prime) \, \frac{dQ}{ds^\prime} ds^\prime \; ,
\\
{\cal I}&=&-\frac{(a+1)(\beta+1)z_0\omega^2}{2g} \int_{1}^{\infty} (s^\prime)^{\mu} \, \psi_{\parallel}(s^\prime) \, \frac{dQ}{ds^\prime} ds^\prime \; ,
\\
\Theta &\equiv& 2 \omega \sqrt{\frac{az_0}{g}\left(1+\frac{\gamma\beta}{2}\right)} \; .
\end{eqnarray}

\noindent The parameter $\Omega$ is a constant to be later determined by matching condition. The function $\psi_{\parallel}$ is the wavefunction for the upward propagating sausage wave. 

The vertical velocity for sausage waves in the isothermal region can be obtained using a finite energy boundary condition applied in the limit of infinite height and by imposing the continuity condition for the vertical component of velocity and the Lagrangian pressure perturbation at the interface with the polytrope. The result is

\begin{equation}
v_\parallel = \left[C e^{-\alpha_{\parallel} (z+z_0)}\ +\ D e^{-\alpha_{d}(z+z_0)}\right]\; ,
\end{equation}

\noindent with the parameters $C$, $D$ and $\Omega$ being known constants that depend on $\alpha_{e}$, $\alpha_{\parallel}$, $\beta$, $\omega$, $a$ and $k_{x}$ (see Gascoyne et al. (2014)). The term involving the decay rate $\alpha_{\parallel}$ is the homogeneous solution for the sausage wave,

\begin{eqnarray}
\label{insidew}
\alpha_{\parallel} = -\frac{1}{4H} + \sqrt{\frac{\omega_s^2 - \omega^2}{c_T^2} } , \hspace{0.5cm}
\omega_s^2 = N^2 + \frac{c_{T}^{2}}{H^{2}}\left(\frac{3}{4}-\frac{1}{\gamma}\right)^{2} ,  \hspace{0.5cm} c_T^2 = \frac{2}{2+\gamma\beta} c_{e}^2 ,
\end{eqnarray}

\noindent and the term with $\alpha_d$ is the inhomogeneous (or driven) 
solution with

\begin{eqnarray}
\alpha_{d} = \frac{1}{4H} + \alpha_e.
\end{eqnarray}

Note that the vertical decay rate of the sausage tube wave and the external $p$ mode differ. Thus, we expect the ratio of power between the internal and external motions to be height dependent. The sausage wave functions depend on the driving frequency of the $p$ modes and the tube's plasma condition described by the parameter $\beta$. Therefore, the power ratio obtained by taking the square of the velocity-amplitudes will also be a function of these two parameters. We investigate this dependence in more details in the next section.

\section{Results}

In Figure 2, we plot 3 mHz $p$-mode eigenfunctions (solid lines) and the corresponding excited sausage wavefunctions (real part with dashes and imaginary part with dotted lines) as functions of depth. The sausage wavefunctions are calculated assuming the plasma $\beta = 0.1$ (left panel) and $\beta = 1$ (right panel) in the flux tube; both the wavefunctions, the $p$ modes and the sausage waves, are normalised with the square root of the mass density such that waves with a spatially uniform energy density will have a constant apparent amplitude. As expected, the amplitudes and wavelengths vary with the plasma $\beta$. Above the interface (the interface is denoted by a vertical dotted lines), the amplitudes appear to be suppressed in low-$\beta$ (high magnetic field strength) flux tubes.

We now compute the acoustic power by squaring the modulus of the velocity wavefunctions $P = |v|^2$. We then define the power ratio $r_n$ for each mode order in isolation as the power $P_{\parallel,n}$ of longitudinal waves inside the tube to the (external) $p$ mode power $P_{e,n}$. Note that such theoretically calculated power ratios are not normalised and thus, cannot be used for direct comparison with observations. Also, the observational
power maps lack wavenumber discrimination. Thus, the measured power ratios are actually a weighted average over mode order of these suppression factors. We shall derive the appropriate weights as follows.

\subsection{Simulating observed power ratios}

Moretti et al. (2007) reported an increase in the spatial distribution of power deficit with height for frequencies less than 5 mHz. Chitta et al. (2011) showed power deficit in $p$-band frequencies for magnetic structures with relatively weaker magnetic field strengths (i.e., $|B| <$ 500 G).  Earlier studies also measured power suppression in plage for $p$ band frequencies where pixels of like field strengths were binned (see for example, Jain and Haber, 2002 and references therein). How can we compare our theoretically obtained power ratios with such variety of observationally measured power maps?

In an unmagnetized pixel of quiet Sun, we model the observed vertical velocity as the the sum over p modes evaluated {\sl at the height of formation} of the spectral line $z_{obs}$,
\begin{eqnarray}
v_{e}(\omega) = \sum_{n}A_{n}(\omega)v_{e,n}(z_{obs};\ \omega),
\end{eqnarray}

\noindent where $A_{n}(\omega)$ is an amplitude that depends on frequency $\omega$ and radial order $n$. 

\noindent In a magnetized pixel the Dopplergram will return an area weighted average of the velocity,
\begin{eqnarray}
v_{mag} (\omega) = \sum_{n}\left[(1-f)A_{n}(\omega)v_{e,n}(z_{obs};\ \omega) + fA_{n}(\omega)v_{\parallel,n}(z_{mag};\ \omega)\right],
\end{eqnarray}

\noindent where $f$ is the filling factor defined as the fractional area occupied by the magnetic fields. Note that the height of formation of the spectral line might be different in a magnetized region, i.e., $z_{mag} \le z_{obs}$.

The observed power is now obtained by taking the square of the modulus of the Dopplergram velocity. If we assume that modes of different order are spectrally isolated and do not interfere, i.e., the frequency separation between modes is much larger than the line widths such that the modes are well-separated in the power spectrum, then the power lacks the mode cross-terms in the product
\begin{eqnarray}
P_{e}(\omega) = \sum_{n}|A_{n}(\omega)|^{2} |v_{e,n}(z_{obs;\ \omega})|^{2},
\end{eqnarray}
\begin{eqnarray}
P_{mag}(\omega) = \sum_{n}|(1-f)A_{n}(\omega)v_{e,n}(z_{obs;\ \omega})\ +\ fA_{n}(\omega)v_{\parallel,n}(z_{obs;\ \omega})|^{2}.
\end{eqnarray}
The first of these two equations is just the summation of the power within each individual mode,
\begin{eqnarray}
P_{e}(\omega) = \sum_{n}P_{e,n}(\omega),
\end{eqnarray}
\noindent while the second equation can be manipulated to reveal that there is a interference term.
\begin{eqnarray}
\label{interference}
P_{mag}(\omega) = \sum_{n}|A_{n}(\omega)|^{2} \left\{(1-f)^{2}|v_{e,n}(z_{obs;\ \omega})|^{2}\ +\ f^{2}|v_{\parallel,n}(z_{mag;\ \omega)}|^{2}\right.\nonumber \\
\left. \ +\ 2f(1-f)Re\left[v_{\parallel,n}(z_{mag;\ \omega})v_{e,n}^{*}(z_{obs;\ \omega})\right]\right\}.
\end{eqnarray}

After factoring out the power in each external mode, we obtain
\begin{eqnarray}
P_{mag} = \sum_{n}P_{e,n}(\omega){\cal P}_{n},
\end{eqnarray}
\noindent where
\begin{eqnarray}
\label{calP}
{\cal P}_{n} = \left[(1-f)^{2} + f^{2}r_{n}\ +\ 2f(1-f)r_{n}^{1/2} \cos (\theta_{\parallel,n} - \theta_{e,n}) \right],
\end{eqnarray}

\noindent and $r_{n} = \frac{|v_{\parallel,n}(z_{mag};\ \omega)|^{2}}{|v_{e,n}(z_{obs};\ \omega)|^{2}}$. In this equation, $\theta_{\parallel,n} - \theta_{e,n}$  is the phase shift between the sausage wave and the external $p$ mode each evaluated at the height of formation appropriate for them,
\begin{eqnarray}
\theta_{\parallel,n} \equiv {\rm arg}\left\{v_{\parallel,n}(z_{mag}; \ \omega)\right\}, \nonumber \\
\theta_{e,n} \equiv {\rm arg}\left\{v_{e,n}(z_{obs};\ \omega)\right\}.
\end{eqnarray}

Thus, the observed power ratio, which is the sum over all radial orders $n$, is
\begin{eqnarray}
\label{sig}
\sigma(\omega) =\sum_{n} \sigma_{n}(\omega)
\end{eqnarray}
\noindent i.e.
\begin{eqnarray}
\label{sigma}
\sigma (\omega) = \sum_{n}p_{n}(\omega)\left[(1-f)^{2} + f^{2} r_{n} + 2f(1-f)r_{n}^{1/2}\cos (\theta_{\parallel,n} - \theta_{e,n}) \right],
\end{eqnarray}

\noindent where
\begin{eqnarray}
\label{peen}
p_{n}(\omega) = \frac{P_{e,n}(\omega)}{P_{e}(\omega)} = \frac{P_{e,n}(\omega)}{\sum_{n}P_{e,n}(\omega)}.
\end{eqnarray}

We could obtain observational estimates of $p_n(\omega)$ from any helioseismic data set. Here, we use the helioseismic technique of ring-analysis as implemented in Greer et al. (2014). The power contained in each mode is calculated as the product of the line-width and amplitude, with these parameters obtained by fitting a Lorentzian function to each mode's power profile. We show such power for different modes in Figure 3. These helioseismic observations are performed using data from the Helioseismic and Magnetic Imager (HMI) on board the Solar Dynamics Observatory (SDO). This instrument uses a spectral line that is formed roughly 200 km above the photosphere (see, Fleck et al. 2011), and therefore the power estimates at $z_{HMI}$, are valid for $z_{HMI} = z_{\rm photo}$+200 km (see Figure 1). Note that in Figure 3, the qualitative behaviour of the $f$ (denoted by black crosses) and for $n>5$ (turquoise squares) modes is quite different compared to the remaining modes. This suggests that the weighting is biased for these modes.

In Figure 4, we plot ${\cal P}_{n}$, as shown in equation (\ref{calP}), as a function of frequency for each mode. We choose $\beta = 1$ and three different filling factors $f = 0.1$ (left), $0.5$ (middle) and $1.0$ (right). The curves are labelled as $z_{obs} = z_{HMI}$. However, recall that the formation heights (FH) of the spectral lines depend on the background atmospheric densities which are different in the magnetic tube compared to the external medium. We approximate the height of formation in a magnetic region as the height where the overlying integrated column mass matches that for the `nonmagnetized' external region. For an isothermal atmosphere, this is equivalent to matching the density in the magnetized region with that in the nonmagnetic region. We thus, consider $z_{mag}$ to be at the same density as $z_{obs}$. In calculating $\sigma_{FH}$, we vary $z_{obs}$ for the $p$ modes and compute $r_{n}$ with $z_{mag}$ corresponding to the height in the magnetic flux tube where the two densities are the same (i.e. where $\rho_{e} = \rho_{i}$). Note from Figure 4 that for low order modes, we have power enhancement ($> 1$) for low frequencies. This is significant for high filling factors. However, these theoretically calculated ${\cal P}_{n}$ are not normalised and thus, cannot be used for direct comparison with observations. We now consider normalisation using the observationally measured power of $f$ and $p$ modes for each frequency.

 In Figure 5, we plot the suppression factor predicted by our model for a variety of photospheric filling factors and values of plasma $\beta$. The solid and dashed black lines show $\sigma$ as a function of frequency computed in two different manners. To illustrate the effect that the magnetic field has by changing the height of formation, as the dashed curves we show the suppression factor for the case where the height of formation doesn't change, i.e., $z_{mag} = z_{obs}$. the colored symbols indicate the contribution made to the observed suppression by each radial mode order. The solid black curve shows the suppression factor when the height of formation is lowered in magnetic regions because of the reduced mass density. Clearly, the power is generally suppressed when the height of formation is lower in the magnetic region.  Both of the simulated curves show that small filling factors produce weak suppression, as expected. It is clear from the figure that the suppression varies with increasing filling factor (left to right panels) and plasma beta (top and bottom panels). Apart from a few isolated data points, generally the suppression factor is a decreasing function of frequency, similar to the observationally measured ratios of this nature. For larger filling factors (which would suggest that observed area is dominated by magnetic fields) the power suppression can vary significantly depending on plasma $\beta$ and frequency.  Thus, two magnetic structures with the same average plasma $\beta$ but different average flux density, can have very different power suppression because of an increase in the area of the magnetic structure that is occupied in the observed field of view. 
 The resulting suppression factor $\sigma_{\rm FH}= \sum_n \sigma_{n,{\rm FH}}(\omega)$
is overplotted as solid line. Clearly, the power is generally suppressed more when the vertical velocities are measured at a height where the densities are same in the magnetised and nonmagnetised regions. 

\section{Discussion}
Theoretically obtained power deficits within strong magnetic fields occur due to a decrease in the skin depth of the sausage waves. The lowering of the height of formation within magnetized regions generates significantly stronger power suppression. Therefore, in all subsequent figures we will only consider the suppression factor $\sigma_{FH}$ where this effect has been accounted for.

It is likely that the power contained in each $p$-mode, itself varies with height and this may influence the power ratio in addition to any change in the filling factor. We now consider this possibility.

\subsection{Variation of the suppression ratio with height}
 To estimate power ratios as they would appear at different heights of observations in the atmosphere we can scale the power measurements as follows:

\begin{eqnarray}
\label{heightP}
p_{n}(z_{obs},\omega) = \frac{p_n(z_{\rm HMI},\omega)}{\sum_{n'} p_{n'}(z_{\rm HMI},\omega) \, \exp\left[-2(\alpha_{e,n} - \alpha_{e,n'})(z_{obs} - z_{\rm HMI})\right]}
\end{eqnarray}

\noindent where $\alpha_{e,n}(\omega)$ is the vertical decay rate of the $p$-mode solution for radial order $n$ in the isothermal atmosphere. Using this expression allows us to estimate the power suppression factor $\sigma(z_{obs}, \omega)$ that would be valid high in the atmosphere (i.e., at $z = z_{obs}$) using acoustic power measurements made with the low-lying HMI spectral line ($z_{\rm HMI} \approx z_{\rm photo} + 200$ km). Note that changing $p_{n}$, according to equation (\ref{peen}) changes the fractional power of each individual mode with height. So for individual modes of a given frequency, the suppression factor decays with the same rate for all values of $\beta$. However, since $r_{n}$ depends on the plasma $\beta$ of the tube, the collective suppression factor $\sigma_{FH}(\omega)$ is expected to decay differently with height for different $\beta$ values.

In our model the filling factor of the magnetic tube will increase with height because the magnetic tubes flare with height. Thus, the changes in filling factor are incorporated as:
\begin{eqnarray}
\label{fill}
f = f_{0}{\rm exp}\left(\frac{z_{obs} - z_{HMI}}{2H}\right).
\end{eqnarray}

Note that Moretti et al. (2007) ascribe mean heights of 100 km, 250 km and 500 km for Ni, K and Na spectral lines for quiet Sun, but with uncertainity over the effect of magnetised plasma on the response function of these spectral lines these values would be only an approximation for the magnetic elements discussed here. Incorporating our findings so far that the power suppression depends on magnetic field strength, filling factor and the formation height of spectral lines, what frequency dependence do we expect for measured power as a function of height in a magnetic element from our model? We therefore, calculate the power suppression factor from equation (\ref{sigma}) with equations (\ref{heightP}) and (\ref{fill}). In Figure 6, we plot this power suppression factor $\sigma_{FH}(z_{obs}, \omega)$ as a function of frequency, for different heights measured above $z=z_{{\rm HMI}}$. These are denoted by different linestyles for two different beta values $\beta = 0.1$ (top panel) and 1.0 (bottom panel). 

Clearly, the rate at which the power suppression varies with height depends on the frequency of the mode and the plasma $\beta$ of the magnetic element and what fractional area this magnetic element occupies at a given height but as found in observationally measured power maps (see Moretti et al., 2007) the power suppression in the same magnetic element increases with height for all modes below the acoustic cut-off frequency (5.2 mHz in our model).

\subsection{Contamination in the external wavefield}
Scattering and mode-mixing by the magnetic fibril causes modifications to the acoustic wavefield in the region external to the tube (e.g., Hindman \& Jain 2012, Gordovskyy et al. 2009). Such effects should manifest as a redistribution of power between the different p-mode orders in the vicinity of a magnetized region. Such a redistribution has not been accounted for here as we assume that the scattering is essentially weak. Further, the external acoustic field in the near-field around a magnetic element should be contaminated by the continuum of acoustic jacket modes that reside on the boundary between magnetic and nonmagnetic regions. However, it is not believed
that the jacket contributes significantly to the observed Doppler velocity power obtained by instruments with the current spatial resolution (Cally 2013). Nevertheless, these issues may be important and are under consideration for future work. In particular, the power suppression from the current theory do not show the same frequency dependence as the observations. This suggests that additional physical effects such as including acoustic jacket modes in the external wavefields etc. need to be explored.

In deriving the formula for theoretical power ratios, it was assumed that the $p$ modes are spectrally isolated and that there is no cross-talk between the line-widths of each individual mode. However any observed $p$-mode power (HMI dataset used here) clearly have finite linewidths and contamination from the adjacent modes which is difficult to isolate. This gives errors in the weighting for each modes. Thus, the computed power suppression may have this error and therefore, requires caution when direct comparisions with the observed power ratios are considered. Other temporal and spatial-specific sources of errors such as pertaining to latitudes, longitudes or magnetic fields are also not well known for the oberved $p$-mode powers. Thus, error bars cannot be estimated in the subsequently computed power suppression factors.

\subsection{The Thin Flux-Tube Approximation}
 
The assumption that the flux tubes are thin ignores lateral variations in the equilibrium and perturbed quantities. Such an assumption is valid provided the local radius, $r_0$ of the tubes is much smaller than any other relevant length scale.  In reality, since magnetic flux tubes flare as a function of height, and the {\sl thin} flux tube approximation must break down at the height where the tube's flux starts to mingle with the flux from neighboring tubes to form a horizontal magnetic canopy. As a related assumption, lateral thermal equilibration ensures that the plasma $\beta$ is constant with height within the flux tubes. This imposes a strong restriction on the height structure of the flux tube. This too breaks down in the chromosphere and a constant $\beta$ model cannot realistically replicate the upper chromosphere. Therefore, caution is required in using our theoretically calculated power ratios when the height of measurement is above 500 km. We take different values of $\beta$ between 0.1 to 1.0 to investigate power ratios (see also, Jain et al. 2011). However, the weak dependence of power suppression for $\beta > 0.5$ (see Figure 6) suggests that lateral variation in the equilibrium quantities of magnetic flux tube may be quite important.

The model density and pressure $\rho_{0}$ and $p_{0}$ at $z_{0}$ are chosen to coincide with the reference model of Maltby et al. (1986) at temperature minimum. Clearly, this temperature is quite low to represent the entire upper atmosphere. However, the details of temperature structure in the upper atmosphere is not that crucial as long as the waves (p mode and sausage mode alike) are evanescent. Thus, all the results were also investigated with higher values of temperature and we found that the qualitative behaviour of the power ratio does not change significantly.

\section{Conclusions}
We investigated the excitation of longitudinal compressible tube wave by solar $p$ modes and studied their velocity amplitudes.
It is found that the interaction of $p$ modes with magnetic field depends on two main competing effects of magnetic field: (1) how easy it is to excite the compressible disturbances in a magnetic flux tube and (2) how rarified is the gas inside the magnetic flux tube for compressible disturbances to propagate efficiently. Since the $p$ modes (acoustic gravity waves) have vertical wavelengths that vary with height, the compressible disturbances excited by them inside a magnetic tube also have varying vertical wavelengths. Thus, amplitudes of the compressible disturbances are modified by the magnetic field. The amplitudes can be broadly compared based on the two regimes characterised by a cut-off frequency. The evansecent regime for frequencies less than the acoustic cut-off frequency, the amplitudes of low order longitudinal compressible waves are smaller in comparison with the external acoustic modes in the isothermal region. However, for frequencies above the cut-off frequency, the waves are no longer trapped waves; they are propagating and strictly speaking, cannot be considered for comparision with observationally measured power ratios of $p$-band frequencies. 

Theoretically calculated power ratios (i.e. the ratio of velocity amplitude squares of sausage wavefunctions to the external $p$ modes) for $p$-band frequencies (below the acoustic cut-off) depend on many factors: (1) reduction in amplitude caused by the velocity difference between the sausage wave and the external $p$ modes (2) a change in the filling factor with height (3) magnetic field strength variation in a magnetic structure (4) height variation in the power contained in the $f$ and $p$ modes.

We conclude from the current study that the power suppression measured in observational data for $p$-mode frequencies in magnetic elements is mainly due to a decrease in the attenuation length of the longitudinal/sausage wave in the magnetic flux tube. More power suppression is found when power ratios are calculated for heights of same densities (this occurs for $z_{mag} < z_{obs}$ in our model) compared to a fixed height $z_{mag} = z_{obs}$ for the same frequency, plasma $\beta$ and filling factor. Thus, we conclude that observed power suppression is quite sensitive to the formation height of the spectral lines; in particular for high filling factors.

\acknowledgments

This work is supported by STFC (UK). BWH also acknowledges NASA grants NNX09AB04G, NNX14AC05G and NNZ14AG05. The data used here are courtesy of NASA/SDO and the HMI science team. SDO is a NASA mission, and the HMI project is supported by NASA contract NAS5-02139.



\begin{figure}
\epsscale{1.1}
\plotone{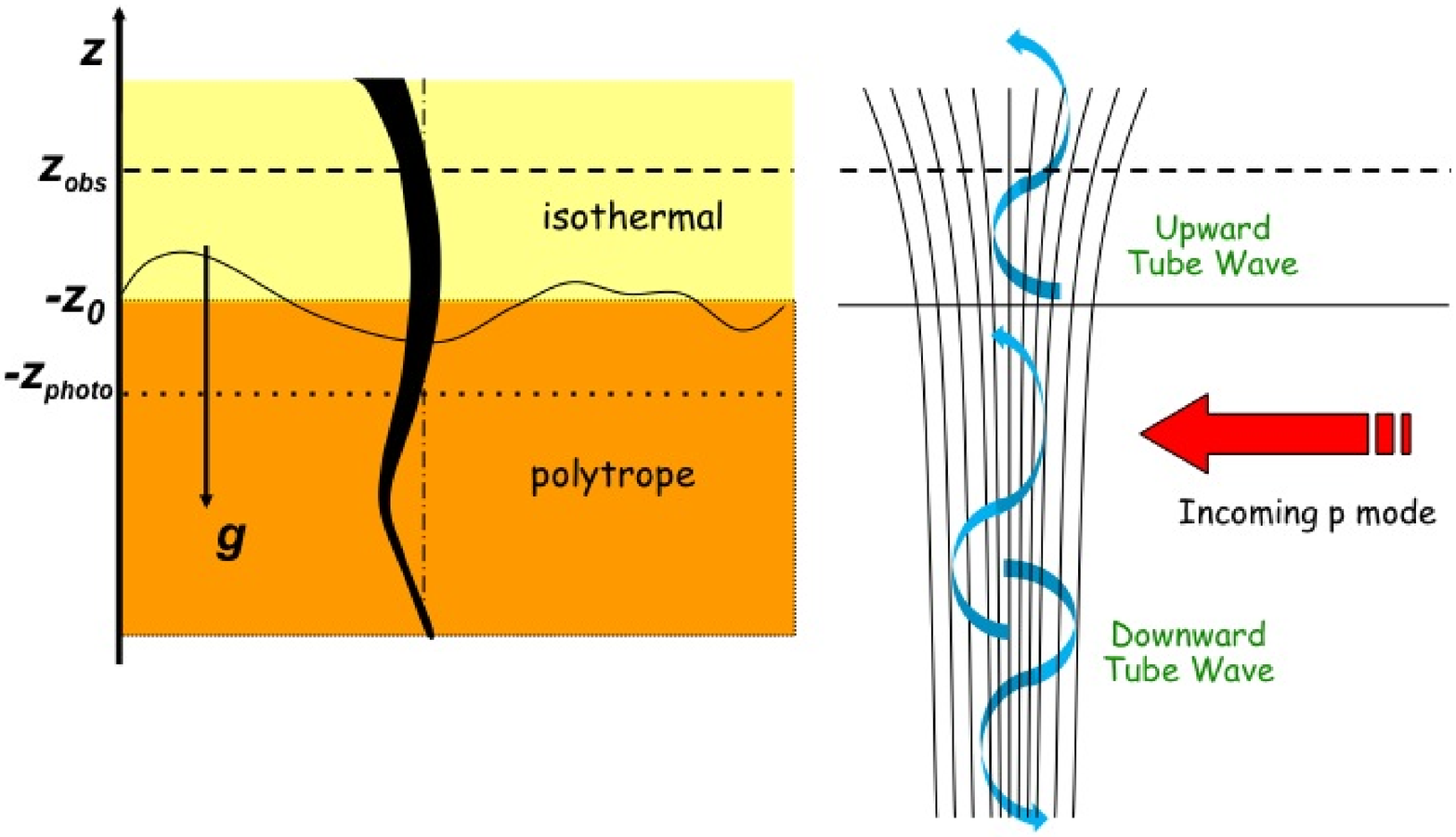}
\caption{Cartoon sketch of the model used. \label{fig1}}
\end{figure}

\begin{figure}
\epsscale{1.0}
\plotone{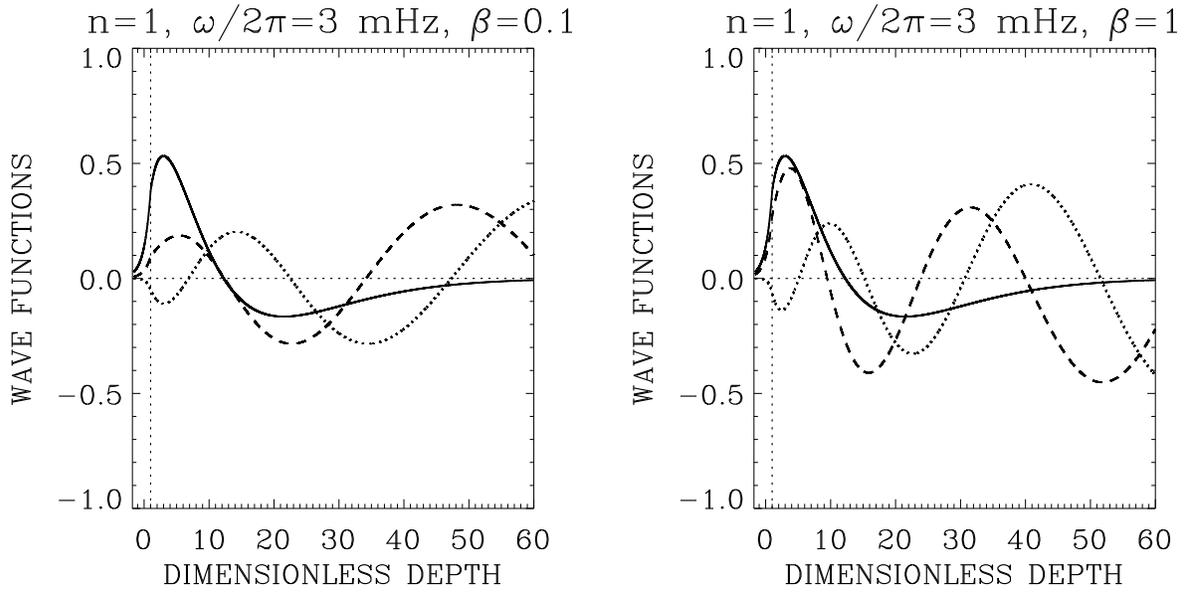}
\caption{The vertical displacement of $p$ modes (solid) and longitudinal waves (real part:dash, imaginary part: dot), normalised by square root of density, as a function of dimensionless depth $s$. The vertical dotted line shows the position of the interface where the polytrope and isothermal regions are matched.\label{fig2}}
\end{figure}

\begin{figure}
\epsscale{0.65}
\plotone{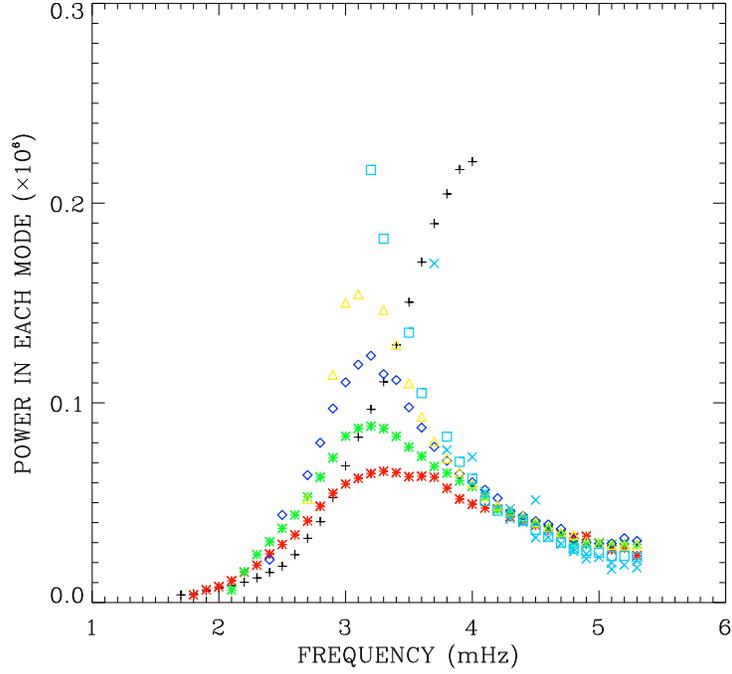}
\caption{Power in each mode as a function of frequency for HMI dataset (used). Different symbol represents a different radial order mode i.e. $n$ = 0 (black crosses), 1 (red asterisks), 2 (green asterisks), 3 (blue diamonds), 4 (yellow triangles), 5, 6, 7 (turqouise). \label{fig3}}
\end{figure}

\begin{figure}
\epsscale{1.0}
\plotone{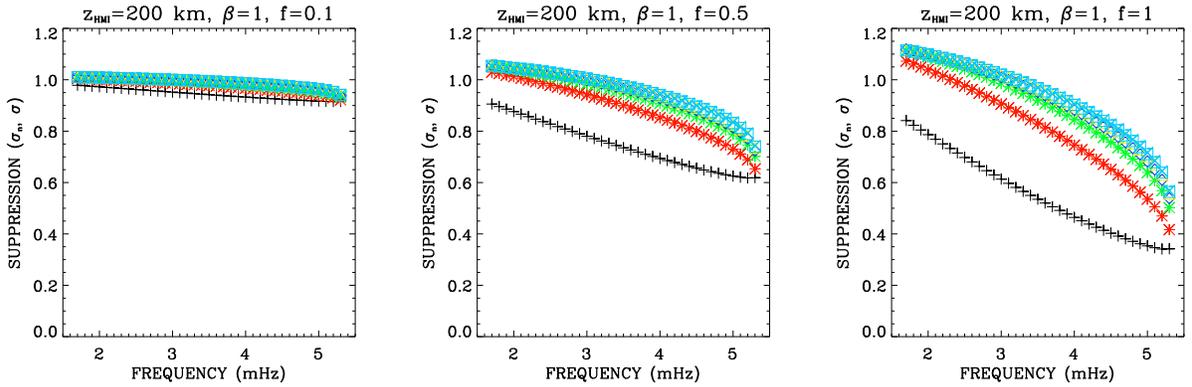}
\caption{${\sigma}_{n}(\omega)$, as shown in equation (\ref{sig}) without the normalising factor ${\cal P}_{n}(\omega)$, as a function of frequency for each radial order (see Figure \ref{fig3} for different symbols). \label{fig4}}
\end{figure}



\begin{figure}
\epsscale{1.0}
\plotone{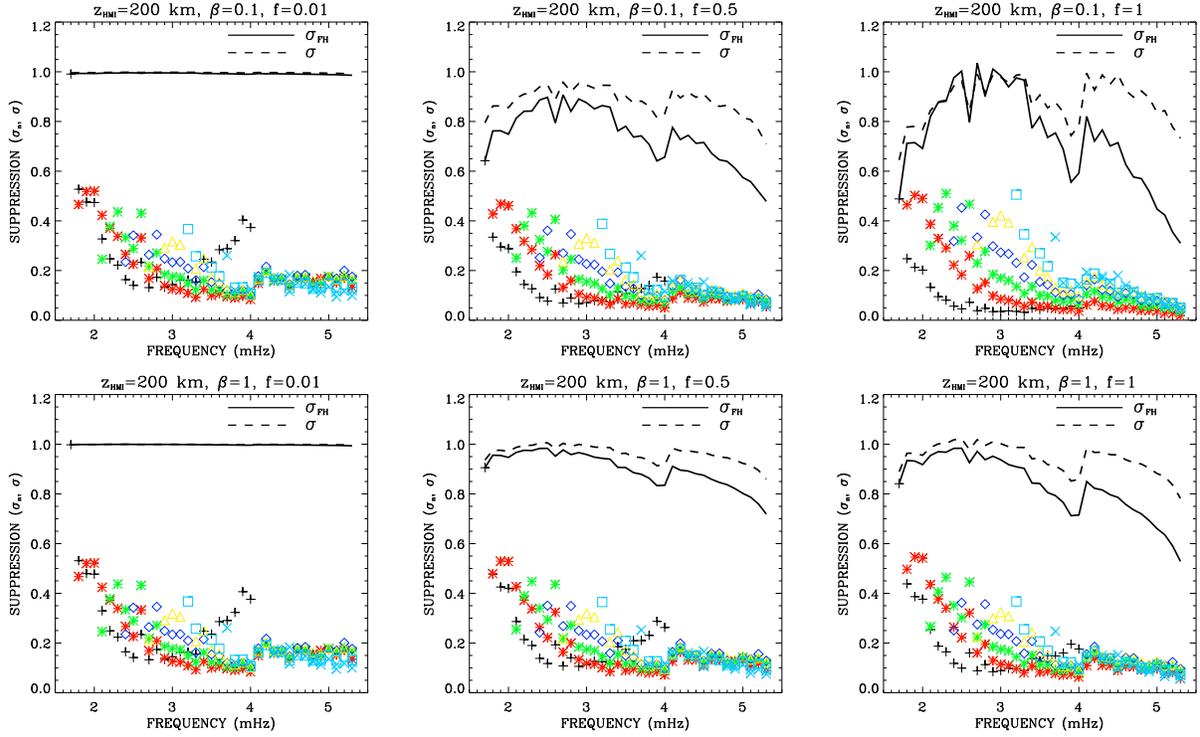}
\caption{The suppression factor $\sigma_{n}$ and $\sigma$ (refer to equation (\ref{sigma}), measured at a height of $z_{obs}=z_{HMI} = 200$ km above the $z_{\rm photo}$, as a function of frequency. Different coloured symbols denote different radial orders $n$ = 0 (black crosses), 1 (red asterisks), 2 (green asterisks), 3 (blue diamonds), 4 (yellow triangles), 5, 6, 7 (turqouise). The tope and bottom panels are for magnetic flux tube with $\beta = 0.1$ and 1.0 respectively. Also, note the three different values of the filling factor $f$.  The dashed line denotes $\sigma = \sum_{n} \sigma_{n}$ for $z_{mag} = z_{obs}$ where as the solid lines ($\sigma_{\rm FH}$) are for when $z_{mag} < z_{obs}$ (i.e for the same $\rho_{i} = \rho_{e}$).\label{fig5}}
\end{figure}

\begin{figure}
\epsscale{0.65}
\plotone{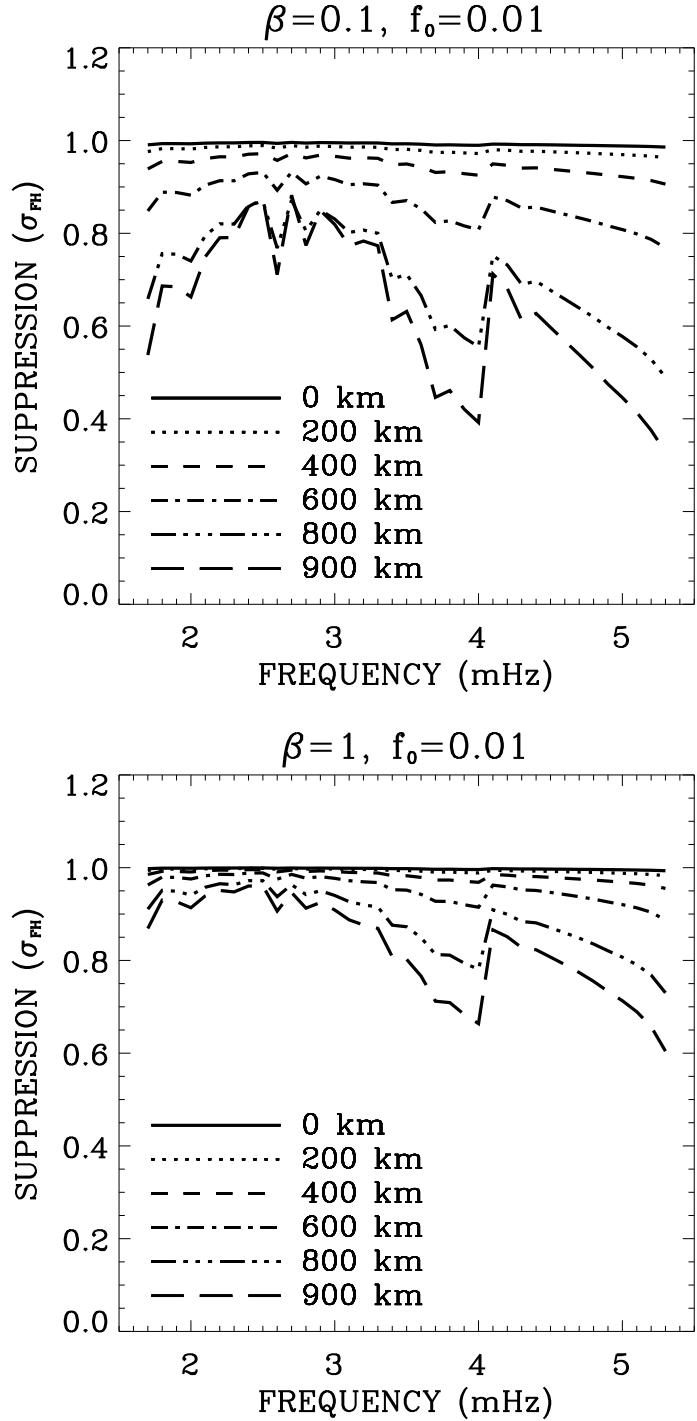}
\caption{The estimated power suppression $\sigma_{FH}$ as a function of frequency.  Different curves denote different height above the photosphere $z$=0 (solid), $200$ km (dot), $400$ km (dash), $600$ km (dot-dash), $800$ km (dot-dot-dot-dash), $900$ km (long dash).  The curves are shown for two different values of beta. \label{fig6}}
\end{figure}




\end{document}